# Effect of Quantum Fluctuations on Magnetic Ordering in CaV$_3$O$_7$[*]

Hiroshi Kontani, M.E. Zhitomirsky[†] and Kazuo Ueda

*Institute for Solid State Physics, University of Tokyo,*
*7-22-1 Roppongi, Minato-ku, Tokyo 106*

(December 24, 2013)

We present a theoretical model for CaV$_3$O$_7$: the 1/4-depleted square spin-1/2 Heisenberg model which includes both the nearest-neighbor coupling ($J$) and the next-nearest-neighbor coupling ($J'$), where $J$ and $J'$ are antiferromagnetic. Recent experiments of the neutron diffraction by Harashina et.al. report the magnetic ordering at low temperatures, which may be called as a stripe phase. It is shown that the observed spin structure is not stable in the classical theory. By employing the modified spin wave theory, we show that the stripe phase is stabilized by the quantum fluctuations for $J'/J > 0.69$. In CaV$_3$O$_7$, the coupling constants are estimated as $J \sim J'$ by comparing the theoretical and experimental results.

There has been great theoretical and experimental interest in the physics of quantum phase transition in two-dimensional spin systems. Recently, the family of calcium-vanadium-oxide, CaV$_n$O$_{2n+1}$ ($n = 2, 3, 4$), is added in the list [1]. These compounds have layered structures, where each layer is a regularly depleted square lattice. Because V$^{4+}$-ion possesses one $d$-electron, CaV$_n$O$_{2n+1}$ is represented by $1/(n+1)$-depleted square spin-1/2 Heisenberg model. Experimentally, magnetic properties of CaV$_n$O$_{2n+1}$ compounds are quite different from one another. A spin gap behavior is observed for $n = 4$, but not for $n = 3$ [1–3]. Peculiar geometry of these systems together with effects of frustration introduced by the next-nearest neighbor coupling result in unusual ground state properties. Recently, intensive theoretical studies were done for the ground state of CaV$_4$O$_9$ [4–10]. It is now established that some frustration introduced by the next-nearest-neighbor coupling is indispensable for the realization of the new type of spin disordered state (plaquette-RVB state) [4–6].

In this letter, we study the ground state spin structure of CaV$_3$O$_7$. According to the recent experiment by Harashina et. al. [2], this system shows a magnetic ordering at low temperatures below $T_\mathrm{N} = 23$ K. The lattice structure (1/4-depleted square lattice) of CaV$_3$O$_7$ together with six-spin unit cell is shown in Fig. 1. The nearest- and next-nearest-neighbor coupling constants are expressed by $J$ and $J'$, respectively. We assume $J, J' > 0$ because both of them are expected to originate from the superexchange through the $p$-orbital of the oxygen [4]. In this letter, we will show that the structure observed experimentally (the stripe phase, Fig. 2(b)) is not a stable state but constitutes a saddle-point in the classical theory. It is the quantum fluctuations which stabilize the stripe phase in the 1/4-depleted square lattice. In many cases the continuous degeneracy existed in the classical solutions is lifted by the quantum fluctuations, order from disorder phenomena [11]. However new feature here is that a classically unstable state is selected.

The Hamiltonian of the 1/4-depleted square Heisenberg model is written as

$$H = J \sum_{\langle i,j \rangle}^{\mathrm{n.n.}} \mathbf{S}_i \cdot \mathbf{S}_j + J' \sum_{\langle i,j \rangle}^{\mathrm{n.n.n.}} \mathbf{S}_i \cdot \mathbf{S}_j \qquad (J, J' > 0), \quad (1)$$

where $\mathbf{S}_i$ represents the $S = 1/2$ spin operator located at $i$-th site. First, we consider the classical energy of the Hamiltonian (1). In Fig. 2, we illustrate some typical spin configurations. Within the six-spin unit cell, the classical ground state has the Néel ordering for $J'/J < 1/4$ and is changed to the fan structure for $J'/J > 1/4$ [12]. In the fan phase, $\theta = \left[\pi - \cos^{-1}(J/4J')\right]/2$ and $\varphi = 3\theta$. In the classical theory the transition between these states is of the second order. Here, we stress that the Néel phase for $J'/J > 1/4$ and the stripe phase for any $J'/J$ are *not* stable, i.e., saddle-point unstable states.

In the following, we will consider the effects of quantum fluctuations by using both the linearized spin wave (LSW) theory [13] and the modified spin wave (MSW) theory [14]. The MSW theory is applicable to the collinear phases, the stripe phase and the Néel phase. We express $S = 1/2$ spin operators by boson operators $a_i$ using the Dyson-Maleev transformation,

$$S_i^z = \frac{1}{2} - a_i^\dagger a_i, \ S_i^+ = (1 - a_i^\dagger a_i)a_i,$$
$$S_i^- = a_i^\dagger \qquad \text{for } \uparrow\text{-spin sites} \qquad (2)$$
$$S_i^z = -\frac{1}{2} + a_i^\dagger a_i, \ S_i^+ = -a_i^\dagger(1 - a_i^\dagger a_i),$$
$$S_i^- = -a_i \qquad \text{for } \downarrow\text{-spin sites.} \qquad (3)$$

Then, the spin Hamiltonian (1) is expressed by the boson operators up to quartic terms. We express the sites which has three (two) nearest neighbors as A(B)-sites. Clearly, the spin reduction due to the quantum fluctuations for A-sites and that for B-sites are in general different. Thus, it

---

is necessary to introduce two kinds of chemical potentials, $\mu_a$ for bosons at A-sites and $\mu_b$ for bosons at B-sites. Now the Hamiltonian (1) is represented as

$$H = \frac{1}{2} \sum_{\langle i,j \rangle \in \mathrm{AF}} J_{ij} \left( a_i^\dagger a_i + a_j^\dagger a_j - a_i^\dagger a_j^\dagger - a_i a_j \right)$$
$$+ \frac{1}{2} \sum_{\langle i,j \rangle \in \mathrm{F}} J_{ij} \left( -a_i^\dagger a_i - a_j^\dagger a_j + a_i^\dagger a_j + a_j^\dagger a_i \right)$$
$$+ \sum_{\langle i,j \rangle \in \mathrm{AF}} J_{ij} \cdot \frac{1}{2} a_i^\dagger \left( a_j^\dagger - a_i \right)^2 a_j$$
$$- \sum_{\langle i,j \rangle \in F(\uparrow,\uparrow)} J_{ij} \cdot \frac{1}{2} a_i^\dagger a_j^\dagger \left( a_i - a_j \right)^2$$
$$- \sum_{\langle i,j \rangle \in F(\downarrow,\downarrow)} J_{ij} \cdot \frac{1}{2} \left( a_i^\dagger - a_j^\dagger \right)^2 a_i a_j$$
$$+ \sum_i \mu_i \left( a_i^\dagger a_i - \frac{1}{2} \right) + E_{\mathrm{classical}}, \quad (4)$$

where AF (F) represents the summation over the classically antiferromagnetic (ferromagnetic) pairs. $J_{ij} = J$ ($J'$) for $\langle i,j \rangle$ of a (next-)nearest-neighbor pair. It is also understood that $\mu_i = \mu_a$ ($\mu_b$) for $i \in$ A-sites (B-sites). As the next step, we reduce Eq. (4) to the quadratic form by means of a mean-field treatment. For the stripe or the Néel phase, three mean fields are introduced.

By diagonalizing the mean-field Hamiltonian by the Bogoliubov transformation in the momentum space, we get the magnon-spectra, $\omega_{\mathbf{k}}^\alpha$, where $\mathbf{k}$ is two-dimensional momentum in the magnetic Brillouin zone and $\alpha$ represents a branch of the spectra, $\omega_{\mathbf{k}}^\alpha \le \omega_{\mathbf{k}}^{\alpha+1}$ ($\alpha = 1, 2, \cdots$). The index $\alpha$ runs from 1 to 3 for the stripe or the Néel phase, and from 1 to 6 for the fan phase, reflecting the difference on symmetry. Needless to say, they should be real and positive for any $\mathbf{k}$ and $\alpha$. At zero temperature, the solution of the MSW theory should satisfy the following relations ;

$$\omega_{\mathbf{k}=0}^{\alpha=1} = \Delta, \quad (5)$$
$$1/2 = \langle a_i^\dagger a_i \rangle_{\mathrm{QF}} + \langle a_i^\dagger a_i \rangle_{\mathrm{BC}} \quad (i \in \mathrm{A,B}), \quad (6)$$

where $\Delta$ is the non-negative constant. In Eq. (6), $\langle \cdots \rangle_{\mathrm{QF}}$ represents the expectation value at zero temperature due to zero-point fluctuations, and $\langle \cdots \rangle_{\mathrm{BC}}$ comes from the Bose condensation of ($\mathbf{k}=0$, $\alpha=1$)-mode, which occurs only in case $\Delta = 0$. The sublattice magnetization for A(B)-site, $M_{A(B)}$, which is derived from the asymptotic form of the spin-spin correlation function, is given by

$$M_{A(B)} = \langle a_i^\dagger a_i \rangle_{\mathrm{BC}} \quad \text{for } i \in \mathrm{A(B)}. \quad (7)$$

In an ordered phase, there is a Goldstone mode by symmetry requirement and thus the corresponding Bose condensation occurs. In this case, the condition

$$\Delta = 0, \quad \langle a_i^\dagger a_i \rangle_{\mathrm{BC}} > 0 \quad (8)$$

should be satisfied. On the other hand, a disordered phase solution has a finite excitation gap $\Delta$ for $\mathbf{k} = 0$, and no Bose condensation exists. Thus, the condition

$$\Delta > 0, \quad \langle a_i^\dagger a_i \rangle_{\mathrm{BC}} = 0 \quad (9)$$

is satisfied in the disordered phase. In the MSW theory we solve self-consistent equations for $\mu_a$, $\mu_b$ and the three mean-fields with the conditions (6) and (8) ] (or (9)). Clearly, at the second-order critical point the conditions $\Delta = 0$ and $\langle a_i^\dagger a_i \rangle_{\mathrm{BC}} = 0$ are satisfied.

Let us start from the case $J = 0$, where the problem reduces to the one-dimensional chain shown in Fig. 3. According to Lieb's theorem [15], the total magnetization of the ground state is rigorously given by $|\langle S_z^{\mathrm{tot}} \rangle| = S \cdot |\# \text{ of A-sites} - \# \text{ of B-sites}| = SN/3$, i.e., the ferrimagnetic phase. Here $N$, which is the total spin number of the chain, is assumed to be a multiple of 3. Table I shows the ground state properties calculated by the LSW, MSW theories and the exact diagonalization combined with finite size scaling up to 24 spins (ED). Note that these three results satisfy the Lieb's theorem. The results by the ED are almost rigorous and relative errors are estimated to be less than $10^{-3}$. It is remarkable that even in the system where two kinds of sublattice magnetizations ($M_A \ne M_B$) coexist, results by the MSW theory are reliable and are much better than those by the LSW theory.

Table I  The ground state properties for $J' = 1$, $J = 0$. $E_g$ is the ground state energy per six spins.

|       | LSW    | MSW    | ED     |
|-------|--------|--------|--------|
| $M_A$ | 0.3476 | 0.3955 | 0.3961 |
| $M_B$ | 0.1951 | 0.2910 | 0.2922 |
| $E_g$ | -2.873 | -2.910 | -2.908 |

For the chain system, there is only one mean field in the MSW theory. It should be mentioned that neither $\mu_a$ nor $\mu_b$ is zero even at zero temperature, $\mu_a = -0.1716J'$ and $\mu_b = 0.4001J'$. It means that the solution obtained by the MSW theory is not a simple mean-field solution. Note that $E_g$ calculated by the MSW theory is slightly lower than the exact $E_g$, which may be due to the lack of the variational principle. For the square Heisenberg model, it is known that the chemical potential tends to zero at zero temperature [14]. Thus in this case the solution by the MSW theory at $T = 0$ is nothing but the mean-field solution.

Now we proceed to the magnetic structure for finite $J'/J$. In the limit of $J' \gg J$, we may use a perturbation theory with respect to $J$. It gives us $\delta E_g = 1.220 J \cos\theta$ per unit cell, where $\theta$ is the angle between the moments on neighboring $J'$-chains. Thus it is seen that the quantum fluctuations favor the state $\theta = \pi$, i.e., the stripe phase. For the stripe phase, the loss of the classical energy is $\sim J^2/J'$ while the gain by the quantum fluctuations is $\sim J$. Therefore the stripe phase is expected for $J'/J \gtrsim 1$. The present 1/4-depleted square lattice $S = 1/2$ Heisenberg model may be one of the examples where the quantum fluctuations favor a collinear phase [11].

Next, we turn to the other limit $J' \ll J$. Because the lattice structure is bipartite in this case, the Néel order is expected. By use of the MSW theory, $E_g = -3.038J$ per unit call, and $M_A = 0.2337$ and $M_B = 0.2493$, respectively. ($\mu_a = 2.112 \times 10^{-2}J$ and $\mu_b = -4.090 \times 10^{-2}J$.)

Finally, we study spin structure for general $J'/J$ and determine the phase diagram. For this purpose, we have to analyze the spin structure which is not stable in the classical sense. Thus the LSW theory is not suitable because it produces unphysical imaginary magnon spectra, reflecting the classical instability. On the other hand, the MSW theory is applicable for such an analysis because the self-consistent treatment of the quantum fluctuations may lead to stabilization of the spin structure which is unstable classically. In the MSW calculations for the stripe or the Néel phase, the three mean fields and the two chemical potentials are determined by the self-consistent equation and a unique solution is obtained. More details of the present treatments will be published elsewhere.

The results of the present study are summarized in Fig. 4 and Fig. 5 :

(i) Stripe Phase : This structure is always unstable in the classical sense. Nonetheless, the quantum fluctuations stabilize this magnetic state. By use of the MSW theory, the solution for the ordered phase is found to be stable for $J'/J > 0.6932$. (In case $J = J'$, $\mu_a = -0.1134J$ and $\mu_b = 0.2607J$.) The recent experiments by neutron diffraction [2] report that the stripe order is realized in $CaV_3O_7$. Note that the quantization axis is along the $x$-direction experimentally [2]. In the present theory direction of the ordered moments is arbitrary because spin-orbit couplings are neglected. By comparing the results of theory and experiment, it is seen that the coupling constants $J' \sim J$ are consistent with $CaV_3O_7$. Thus, we conclude that the classically unstable (not a metastable) magnetic structure is stabilized by the quantum fluctuations. This is the unique feature of $CaV_3O_7$.

(ii) Néel Phase : By use of the MSW theory, the solution for the ordered phase is obtained for $0 < J'/J < 0.4437$. Note that the quantum fluctuations stabilize the Néel order beyond the classical critical point, $J'/J = 1/4$. The order-disorder transition at $J'/J = 0.4437$ is of the second order.

(iii) Disordered Phase : The solution for the disordered phase is found for $0.4437 < J'/J < 0.6932$. The minimum of the spin gap, which is shown in Fig. 5, is always at $\mathbf{k} = (0,0)$. The transition between the disordered phase and the stripe phase is of the first-order.

(iv) Fan Phase : The LSW theory predicts that the energy of this phase is always higher than the others except for the very narrow range around $\sim 0.7$. (i.e., $0.6769 < J'/J < 0.7346$.) However, more importantly, it should be noted that the fan phase cannot be stabilized by quantum fluctuations because its spin reduction is always divergent. The energy for this phase obtained by the LSW theory may be unreliable because of lack of self-consistency.

In conclusion we have shown that the magnetic structure of $CaV_3O_7$, the stripe phase, is not stable in the classical theory and that only by taking quantum fluctuations into account stability of the stripe phase is explained. For this purpose we have extended the MSW theory to the system where the reduction of the magnetic moments depends on sites. Finally we would like to emphasize that the regular depletion of the square lattice spin-1/2 Heisenberg model realized in $CaV_nO_{2n+1}$ leads to novel phenomena, the plaquette RVB phase for $n = 4$, and the classically unstable stripe phase for $n = 3$. It is an interesting future problem to elucidate the ground state of the third system, $n = 2$, of this series.

## ACKNOWLEDGMENTS

We would like to thank Masatoshi Sato, Hiroshi Yasuoka and Matthias Troyer for helpful discussions. This work is financially supported by a Grant-in-Aid for Scientific Resarch on Priority Areas from the Ministry of Education, Science and Culture.

fluctuations do not favor such a complicated non-collinear phase.

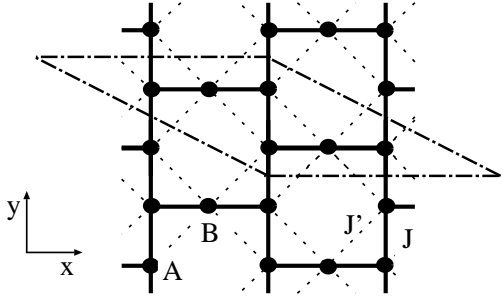

FIG. 1. A model for $CaV_3O_7$ of the spin-1/2 Heisenberg model with the nearest neighbor (solid lines) and the next nearest neighbor (broken lines) exchange interactions. The dot dashed lines show the six-spin unit cell of the 1/4 depleted square lattice.

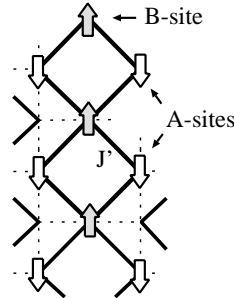

FIG. 3. The chain Heisenberg model constructed only by the next-nearest-neighbor coupling $J'$.

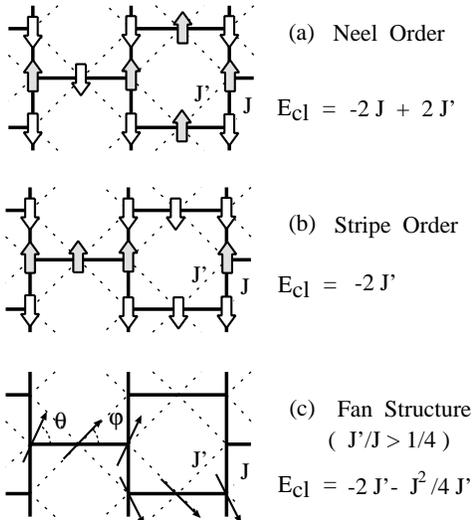

FIG. 2. Typical classical magnetic structures. The fan phase is the ground state for $J'/J > 1/4$ and the Néel phase is the ground state for smaller frustrations. In the classical treatment, the stripe phase is a saddle point and thus always unstable.

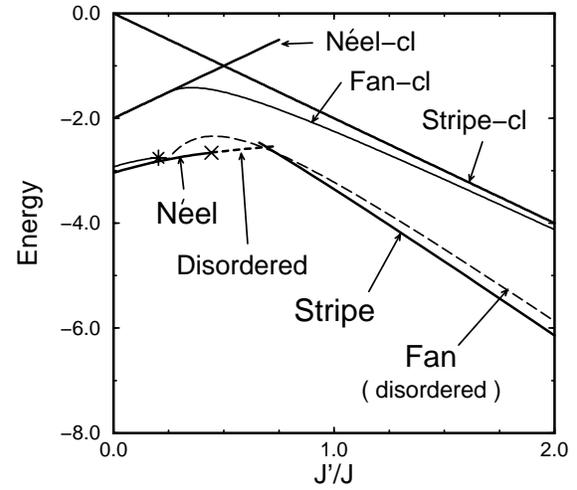

FIG. 4. Comparison between energies of different states. In this figure, 'Stripe-cl', 'Néel-cl' and 'Fan-cl' represent the classical energies of the corresponding magnetic orderings. 'Stripe' and 'Néel' represent the energy calculated by the MSW theory, respectively. 'Fan' represents the energy calculated by the LSW theory, whose spin reduction however is always divergent. Energy of the disordered state is shown by the thick dashed line.

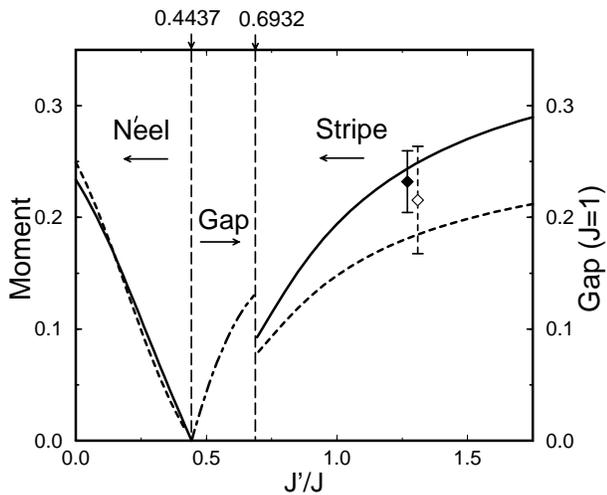

FIG. 5. Phase diagram obtained by the MSW theory, considering the Néel, stripe and the disordered phases. Solid (broken) lines represent the sublattice magnetization for A(B)-sites, respectively. For the disordered phase the magnitude of the spin gap is shown by the dot-dashed line. The experimental data by the neutron diffraction[2] are also shown, where the form factor of $d_{xz}$ (or $d_{yz}$) is assumed.